%% file: paper.tex
\documentclass[sigconf]{acmart}

\makeatletter
\def\@ACM@checkaffil{
    \if@ACM@instpresent\else
    \ClassWarningNoLine{\@classname}{No institution present for an affiliation}%
    \fi
    \if@ACM@citypresent\else
    \ClassWarningNoLine{\@classname}{No city present for an affiliation}%
    \fi
    \if@ACM@countrypresent\else
        \ClassWarningNoLine{\@classname}{No country present for an affiliation}%
    \fi
}
\makeatother

\usepackage{microtype}
\usepackage{graphicx}
\usepackage{subfigure}
\usepackage{booktabs} 
\usepackage{algorithm}
\usepackage{algorithmic}

\usepackage{blindtext}
\usepackage{placeins}

\AtBeginDocument{%
  \providecommand\BibTeX{{%
    \normalfont B\kern-0.5em{\scshape i\kern-0.25em b}\kern-0.8em\TeX}}}

\copyrightyear{2023}
\acmYear{2023}
\setcopyright{acmlicensed}
\acmConference[ICPP 2023]{52nd International Conference on Parallel Processing}{August 7--10, 2023}{Salt Lake City, UT, USA}
\acmBooktitle{52nd International Conference on Parallel Processing (ICPP 2023), August 7--10, 2023, Salt Lake City, UT, USA}
\acmPrice{15.00}
\acmDOI{10.1145/3605573.3605650}
\acmISBN{979-8-4007-0843-5/23/08}

\begin{document}

\title{OSP: Boosting Distributed Model Training with 2-stage Synchronization}


\input{Extra/Author}

\renewcommand{\shortauthors}{Zixuan Chen et al.}


\input{Extra/Abstract}





\maketitle

\input{body}

\FloatBarrier

\bibliographystyle{ACM-Reference-Format}
\bibliography{paper}


\end{document}

%% file: Extra/Author.tex
\author{Zixuan Chen}
\authornote{Equal contribution.}
\authornote{© {Owner/Author | ACM} {2023}. This is the author's version of the work. It is posted here for your personal use. Not for redistribution. The definitive Version of Record will be published in {ICPP 2023}.}
\email{zxchen20@fudan.edu.cn}
\orcid{0000-0002-0126-0387}
\affiliation{%
  \institution{Fudan University}
  \streetaddress{No.220, Handan Rd, Yangpu District}
}
\author{Lei Shi}
\authornotemark[1]
\email{l_shi21@m.fudan.edu.cn}
\orcid{0009-0004-0649-5443}
\affiliation{%
  \institution{Fudan University}
}
\author{Xuandong Liu}
\email{liuxd22@m.fudan.edu.cn}
\orcid{0009-0002-6521-6200}
\affiliation{%
  \institution{Fudan University}
}
\author{Jiahui Li}
\email{jhli22@m.fudan.edu.cn}
\orcid{0000-0002-6559-188X}
\affiliation{%
  \institution{Fudan University}
}
\author{Sen Liu}
\email{senliu@fudan.edu.cn}
\orcid{0000-0003-2230-7671}
\affiliation{%
  \institution{Fudan University}
}
\author{Yang Xu}
\authornote{Corresponding author.}
\email{xuy@fudan.edu.cn}
\orcid{0000-0002-0958-8547}
\affiliation{%
  \institution{Fudan University}
}
\affiliation{%
  \institution{Peng Cheng Laboratory}
}

%% file: Extra/Abstract.tex
\begin{abstract}
    
Distributed deep learning (DDL) is a promising research area, which aims to increase the efficiency of training deep learning tasks with large size of datasets and models. As the computation capability of DDL nodes continues to increase, the network connection between nodes is becoming a major bottleneck. Various methods of gradient compression and improved model synchronization have been proposed to address this bottleneck in Parameter-Server-based DDL. However, these two types of methods can result in accuracy loss due to discarded gradients and have limited enhancement on the throughput of model synchronization, respectively. To address these challenges, we propose a new model synchronization method named Overlapped Synchronization Parallel (OSP), which achieves efficient communication with a 2-stage synchronization approach and uses Local-Gradient-based Parameter correction (LGP) to avoid accuracy loss caused by stale parameters. The prototype of OSP has been implemented using PyTorch and evaluated on commonly used deep learning models and datasets with a 9-node testbed. Evaluation results show that OSP can achieve up to 50\% improvement in throughput without accuracy loss compared to popular synchronization models.

\end{abstract}

%% file: body.tex
\newcommand{\solution}{OSP}

\section{Introduction}

Deep learning (DL) is often used to solve practical problems such as natural language processing (NLP)~\cite{openai2023gpt4}, recommendation systems~\cite{zhang2019deep, DBLP:conf/ijcai/WangHW0SOC0Y21}, online translation~\cite{raffel2020exploring, vaswani2017attention, stahlberg2020neural}, computer vision~\cite{resnet-he2016deep, niemeyer2021giraffe, DBLP:conf/iclr/DosovitskiyB0WZ21}, and more. With the increasing size of datasets and models, the need for more efficient training methods has become essential. Distributed deep learning (DDL) has emerged as a hot field of research and is becoming mainstream to improve training efficiency and meet the demands of larger datasets and models.

New challenges have emerged in DDL training as hardware improves. In the past, bottlenecks in distributed training were primarily on the GPU. However, as GPU performance has increased dramatically, the network has become a major bottleneck. For instance, when using 8 single-GPU computing workers connected to 10Gbps links to train the ResNet152 model with the CIFAR-10 dataset, the communication overhead when using RTX2080Ti (13.45 TFLOPS) is 10\%. However, the communication overhead increases dramatically to 39\% when the GPUs are replaced by RTX3090 (35.58 TFLOPS). Therefore, communication optimization becomes a crucial area of research in DDL~\cite{hsp-li2022hsp}.


The architecture of DDL can significantly affect communication efficiency. At present, the commonly used DDL architectures are Ring-AllReduce~\cite{ring-allreduce} and Parameter Server (PS)~\cite{byteps-jiang2020unified}. The Ring-AllReduce~\cite{ring-allreduce} training architecture aims to minimize communication bottlenecks during synchronization. However, it suffers from severe fault tolerant issues even when the model or number of training nodes is not quite large, resulting in unacceptable training time~\cite{lao2021atp}. On the contrary, PS is not only resilient to fault but also has low implementation difficulties, especially in large clusters.

In PS architecture, synchronization models play a crucial role in the throughput and convergence of DDL~\cite{shi2022robust, aviv2021asynchronous}. Traditional synchronization models used in PS include Bulk Synchronization Parallel (BSP)~\cite{bsp-gerbessiotis1994direct} and Asynchronization Parallel (ASP)~\cite{asp-agarwal2011distributed}. BSP synchronizes the DL model globally at the end of each iteration with barriers, which can lead to the ``incast problem'' aggravating network bottlenecks~\cite{zats2012detail, chen2023boosting}. ASP allows each node to synchronize independently, with a better throughput but worse top-1 accuracy. Stale Synchronous Parallel (SSP)~\cite{ssp-ho2013more} aims to synchronize the global model by controlling cumulative gradient differences but still suffers from accuracy loss and ``stragglers''.

Optimizing the synchronization model has been a challenge and a vital area of research in recent years. Several methods have been proposed to address the issues of network bottlenecks and accuracy loss. R$^2$SP~\cite{r2sp-chen2019round} and HSP~\cite{hsp-li2022hsp} are incremental approaches to SSP, incorporating accuracy correction and batch size tuning techniques to reduce final accuracy loss while improving throughput. However, these methods still have room for improvement.

To summarize the above research, existing approaches to synchronizing distributed deep learning models raise new problems. These methods may still have potential accuracy degradation issues. Moreover, there is still room for communication improvement. We propose a new synchronization model named Overlapped Synchronization Parallel (\solution) to address these challenges. This synchronization model aims to overlap communication and computation more efficiently and enhances throughput through a 2-stage synchronization process. \solution\ splits the synchronization process into Routine Synchronization (RS) for synchronizing important gradients and In-Computation Synchronization (ICS) for synchronizing non-important gradients. By allowing ICS to be overlapped with computation, overall communication time is optimized. While some impact on final accuracy is inevitable due to stale parameters\footnote{Stale parameters refer to outdated values in the deep neural model, which may cause inaccuracies or slow down the training process.}, this can be mitigated by the accuracy correction algorithm called Local-Gradient-based Parameter correction (LGP) included in \solution. 

In summary, the contributions of this paper are as follows:

\begin{enumerate}
    \item We propose \solution, a novel and efficient synchronization model that improves the throughput of DDL training by overlapping communication and computation. This model combines the importance of parameters with the synchronization process for the first time to boost the throughput of distributing model training. 
    \item We propose a refined algorithm to determine the number of parameters to be overlapped with calculations, which does not introduce additional computation overhead for the workers. We provide mathematical proof of the algorithm for ranking the gradient importance and design a parameter correction algorithm to address the accuracy loss caused by delayed synchronization.
    \item We implement a prototype of \solution\ using PyTorch~\cite{paszke2019pytorch} and conduct extensive experiments on a 9-node real testbed under many prevalent DL models including image classification and NLP. The results show that the system using \solution\ can achieve up to 50\% throughput improvement over popular synchronization approaches (ASP, BSP, and R$^2$SP) with no loss of accuracy. 
\end{enumerate}

The rest of the paper is organized as follows. The background and motivation of this paper are described in \S~\ref{section:motivation}. The design concept of \solution\ is introduced in \S~\ref{section:concept}, while the design and implementation details are presented in \S~\ref{section:design}. \S~\ref{section:evaluation} presents extensive evaluation results. The discussions are conducted in \S~\ref{section:discussion} and related works are summarized in \S~\ref{section:related_work}. We conclude the paper in \S~\ref{section:conclusion}.

\section{Background and Motivation} \label{section:motivation}

This section provides an overview of the background of DDL, the PS, and synchronization models. We also demonstrate that the network is a crucial bottleneck in improving the throughput of state-of-the-art deep learning systems with empirical studies. Numerous research efforts have been aimed at optimizing network communication or synchronization models to improve the throughput of DDL training, which we briefly review at the end of this section.

\subsection{Overview of Distributed Deep Learning Training}

\subsubsection{Distributed Deep Learning Training}

The primary goal of DL is to train deep neural networks (DNN) to perform tasks such as classifying input objects or generating specific outputs by adjusting the parameters $\omega$. Given a dataset containing $N$ samples, where the i-th sample is represented by a matrix $x_i$ and has an associated label $y_i$, the objective of the deep neural network is to find the optimal set of parameters $\omega$ that minimize the difference between the predicted output and the accurate output for all samples in the dataset.

\vspace{-0.2in}
$$
Minimize \sum_{i=1}^{N}{loss(f(w, x_i), y_i)}
$$




In data-parallel DDL, multiple workers train on different subsets of the data and perform gradient or parameter synchronization operations with other workers. A PS-based DDL system utilizes one or more PS(s) for synchronization. Each worker will send its computed gradients to the PS, which will then aggregate the gradients by weighting them by the ratio of the size of the worker's subset of the data to the entire dataset. The PS will then merge the gradients to update the global parameters $\omega_g$ and send it back to workers.

\subsubsection{Synchronization Models in PS-based DDL}

\begin{figure}[tbp]
\centering
\begin{minipage}[t]{\columnwidth}
\includegraphics[width=\columnwidth]{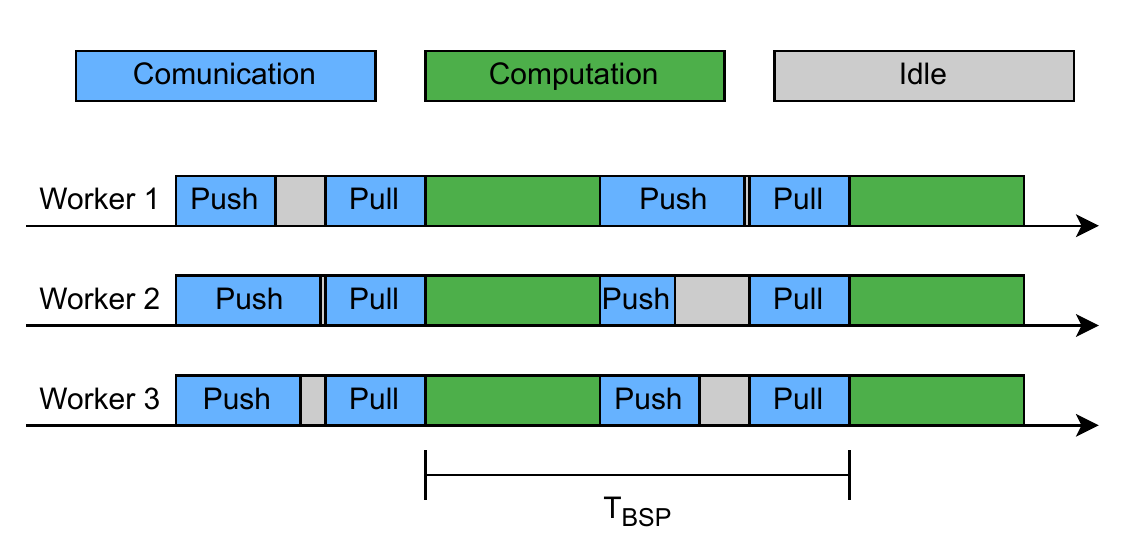}
\vspace{-0.3in}
\caption{Simultaneous communications of the BSP.}
\label{figure:bsp_introduction}
\end{minipage}
\begin{minipage}[t]{\columnwidth}
\includegraphics[width=\columnwidth]{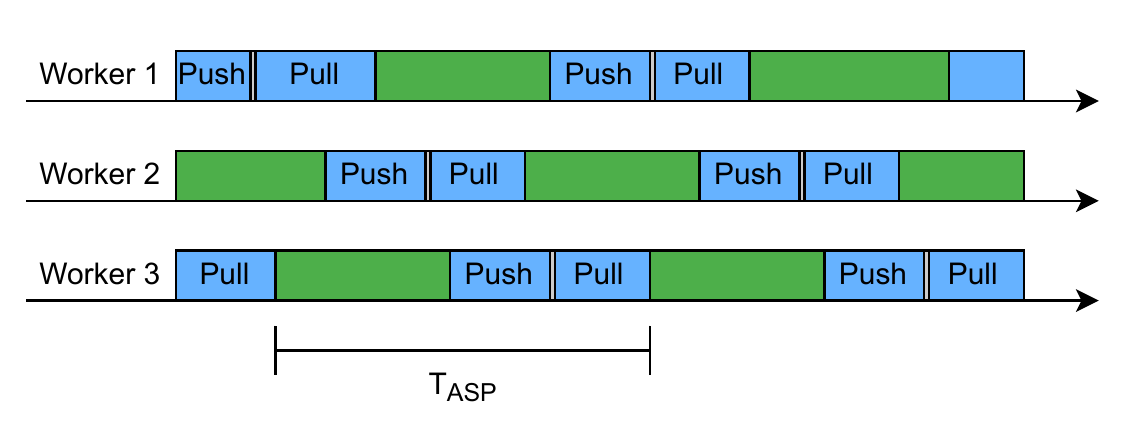}
\vspace{-0.3in}
\caption{Asynchronous communications of the ASP.}
\label{figure:asp_introduction}
\end{minipage}
\end{figure}

In a PS-based DDL training system, synchronizing the neural network parameters among workers is a significant research problem. Popular synchronization models such as BSP~\cite{bsp-gerbessiotis1994direct} and ASP~\cite{asp-agarwal2011distributed} are widely used in DDL training. In BSP, all workers synchronize the global parameters or gradients with the PS after completing the forward and backward propagation (FP and BP) computations. However, this process can be slowed down by the incast problem~\cite{zats2012detail, chen2023boosting} and straggler workers~\cite{r2sp-chen2019round} (as shown in Figure~\ref{figure:bsp_introduction}). In contrast, ASP does not ensure that all workers have the same parameters (as shown in Figure~\ref{figure:asp_introduction}). In ASP, each worker synchronizes its parameters independently with the PS. After each worker pushes its local model parameters to the PS, the PS merges the models into the global model and returns the updated parameters.

We measure the time consumption of each iteration ($T_{BSP}$ in Figure~\ref{figure:bsp_introduction} and $T_{ASP}$ in Figure~\ref{figure:asp_introduction}). Empirical studies have shown that $T_{ASP}$ tends to be up to 6 times smaller than $T_{BSP}$ due to the severe incast and straggler problem in BSP~\cite{syncswitch-li2021sync}, which implies a higher throughput for DDL training with ASP. However, ASP may suffer from significant accuracy loss as it does not maintain global synchronization, and workers may be trained with stale values. 

\subsection{Boosting DDL by Efficient Communication}

\begin{figure}[tbp]
\centering
\includegraphics[width=\columnwidth]{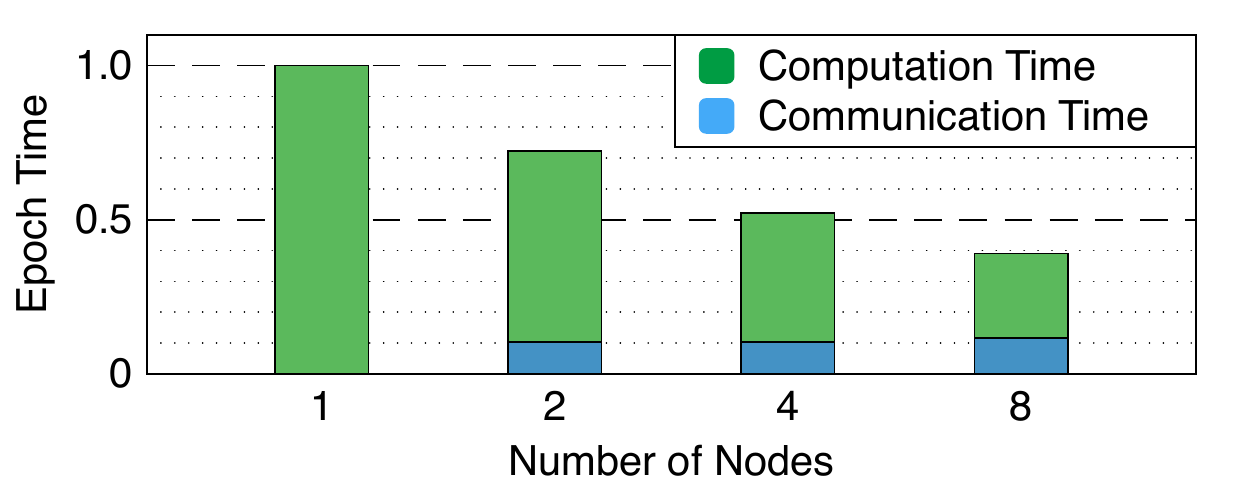}
\caption{Communications share is expanding as DDL training scales.}
\vspace{-0.1in}
\label{figure:communication_share_as_worker_scale}
\end{figure}

It has been observed that communication has become the primary bottleneck in state-of-the-art DDL training systems~\cite{chen2023boosting}, which is primarily due to: 1) the significant increase in GPU/TPU performance and 2) the increased overhead of the network when adding more workers. We evaluate the ResNet50~\cite{resnet-he2016deep} model for PS-based DDL training on 1, 2, 4, and 8 machines (as shown in Figure~\ref{figure:communication_share_as_worker_scale}). The results show that the number of nodes does not proportionally optimize the overall training time, resulting in low cost-effectiveness when scaling the number of workers.

\subsubsection{Synchronization Model Optimization}

Optimizing the synchronization model is an effective solution to improve the throughput of DDL training. Wait-free backpropagation (WFBP)~\cite{wfbp-shi2019mg} uses a directed acyclic graph (DAG) based on the principles of forward and backward propagation in neural networks to overlap communication and computation seamlessly. However, it has high programming limitations and requires significant modifications to the deep learning frameworks. As both BSP and ASP have their advantages and disadvantages, Sync-Switch~\cite{syncswitch-li2021sync} argues that using ASP with stale values in the early training stage can lead to the neural network getting stuck in a local optimum, causing a decrease in final accuracy, while this problem is less prevalent in the later stages of training. To improve training throughput while maintaining final accuracy, they propose using BSP in the early training stage and ASP in the later stage. However, determining when to switch from BSP to ASP can take time and effort.

Other works aim to optimize ASP or BSP directly, like reducing the accuracy loss of ASP or improving the throughput of BSP. SSP~\cite{ssp-ho2013more} artificially sets a specific iteration gap between the fastest and slowest workers to prevent straggler workers from training with overly stale values in ASP. R$^2$SP~\cite{r2sp-chen2019round} uses a round-robin approach for worker-to-PS synchronization to utilize the bandwidth of the PS's duplex links fully. HSP~\cite{hsp-li2022hsp} builds upon R$^2$SP and provides an accuracy correction algorithm for training with stale parameter values for some workers. 

All these works have achieved significant improvements in DDL training throughput. Neverlessness, R$^2$SP still faces the issue of stale parameters when the number of workers grows, and HSP needs to address the problem of non-compliance with training on non-independent identically distributed datasets. None of these attempts get to the crucial point of the synchronization model optimization problem: reducing the amount of data to be synchronized for each iteration, which is the main challenge in this paper. 

\subsubsection{Reducing the Size of Communication}

Research on synchronization models has kept the communication size the same during each iteration. However, many studies have focused on improving communication efficiency by minimizing the amount of communication per iteration. We review state-of-the-art research on gradient compression.

There are two general approaches to reducing the communication size: gradient/parameter quantization and compression. Quantization is a method of representing floating-point numbers using fewer bits. 8-bit quantization converts a 32-bit floating-point number into 8 bits, reducing the size of communication. However, the level of optimization achieved is quite limited. Gradient compression, achieved through sparsification, discards unimportant gradients or parameters during synchronization. For example, the widely used Top-K~\cite{topk-aji2017sparse} and Random-K~\cite{randk-stich2018sparsified} algorithms select the K\% gradients with the highest absolute value or randomly select K\% gradients for synchronization, respectively, to reduce the communication size. These methods have been proven to be mature and have been implemented in several DDL systems.

In addition to the two basic gradient compression methods, several studies have employed more advanced techniques to further reduce communication overhead. For instance, Deep Gradient Compression~\cite{lin2018deep} achieves significant gradient compression by utilizing techniques such as momentum correction~\cite{qian1999momentum} and warm-up training~\cite{goyal2017accurate} to maintain final accuracy.

While all of these methods can alleviate the impact of communication bottlenecks in DDL training, a significant issue is that the loss of gradients or parameters can lead to a significant degradation in accuracy, up to 20\%~\cite{xu2021grace}. Furthermore, overly complex compression algorithms can add computational overhead and may even negate the improvement in communication time, making them cost-ineffective. \solution\ offers a better solution to reduce communication size without introducing additional computation and ensures that no gradients are dropped during DDL training.

\section{\solution\ Design Concept and Challenge} \label{section:concept}

After reviewing the above methods for improving communication efficiency, we conclude that each solution has its limitations. In research on reducing communication size, these methods focus on improving throughput without adequately considering the final accuracy. Additionally, among recent approaches to improve synchronization models, they address the long-tail latency problem~\cite{zats2012detail, chen2023boosting} caused by incast but do not significantly reduce communication time. This poses the fundamental challenge: \textbf{Can we design an approach that combines the benefits of the aforementioned methods?}

\subsection{Overview of \solution}

Designed to further improve training throughput and accuracy, \solution\ ingeniously combines gradient compression with the synchronization model. 

The core concept of \solution\ is to split a portion of the less important gradients and overlap them with the computation, which can remarkably reduce synchronization time and enhance training throughput.

\begin{figure}[tbp] 
\centering
\includegraphics[width=\columnwidth]{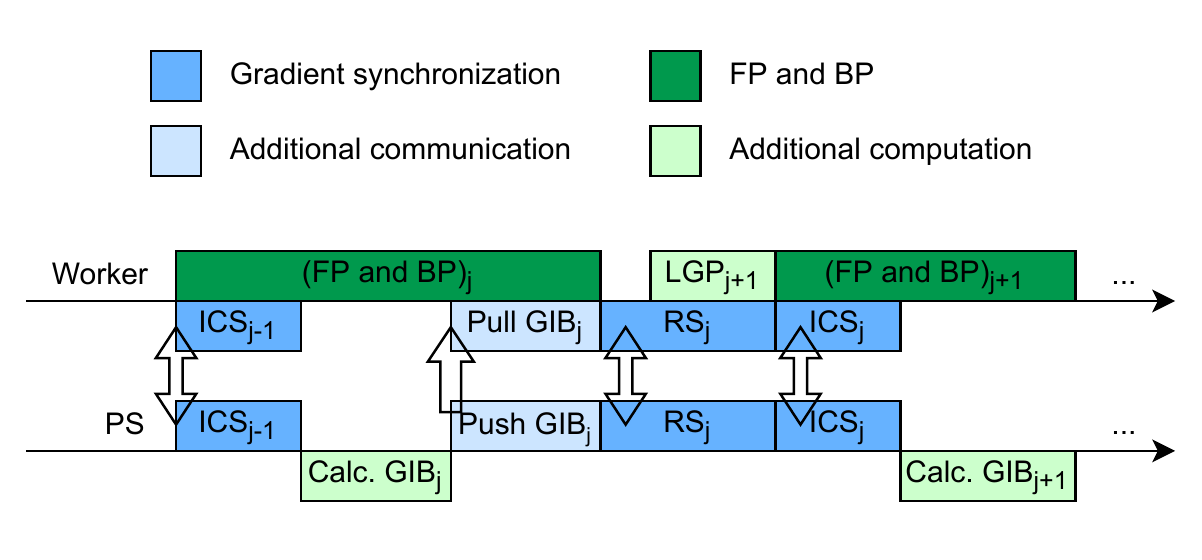}
\caption{\solution\ splits parts of the gradient synchronization to the next iteration of computation to improve the training throughput and adds other steps that do not affect the efficiency of the computation to preserve accuracy.}
\vspace{-0.1in}
\label{figure:OSP_seq_design}
\end{figure}

The synchronization process of \solution\ is a two-stage process, referred to as ``Routine Synchronization (RS)'' and ``In-Computation Synchronization (ICS)'' (See Figure~\ref{figure:OSP_seq_design}). After the computation of j-th iteration, all workers synchronize the important gradients ($G^i$) to the PS during RS while synchronizing the unimportant gradients ($G^u$) at ICS during the next iteration's computation. The workers commence computation immediately after RS, utilizing the prediction of the global gradients based on local gradients. This two-stage synchronization is the key to increasing training throughput.

\subsection{Designing \solution: Challenges and Solutions}

In the design of \solution, several critical challenges must be addressed:

A pressing issue is \textbf{how to distinguish important and unimportant gradients without additional computational overhead} (the first challenge). Most previous approaches for reducing communication size compress gradients after the current iteration's computation, which temporarily halts synchronization. \solution\ employs an alternative strategy: determining the importance of the parameters to rank the importance of the corresponding gradients. \solution\ calculates the importance ranking for the i-th iteration using \textbf{Parameter-Gradient Production (PGP)}, which enables the PS(s) to calculate gradient importance in advance with the worker's computation and return the \textbf{Gradient Importance Bitmap (GIB)} in time, thereby informing the workers which gradients should be synchronized during RS. The gradient importance ranking algorithm is discussed in \S~\ref{sssection:gradient_importance}.

The second challenge that must be addressed is \textbf{determining the number of important gradients to be transmitted during RS.} We propose an algorithm that takes into account the network's link qualities~\footnote{bandwidth, round-trip time (RTT), and loss rate}, computation time per iteration, and experiences from different models. The algorithm for determining the number of important gradients is discussed in \S~\ref{sssection:important_gradient_size}.

The third challenge, which is a significant problem in previous works for reducing the size of synchronization gradients, is \textbf{preserving final accuracy.} Unlike other gradient compression methods, \solution\ does not discard unimportant gradients. Instead, \solution\ synchronizes unimportant gradients during computation, thereby overlapping communication and computation. This approach may raise a new computation issue on stale parameters, which is resolved by predicting global unimportant gradients based on local unimportant gradients and is discussed in more detail in \S~\ref{sssubsection:preserve_accuracy}.

\section{\solution\ Design Details} \label{section:design}

In continuation with the previous section, we present the comprehensive design specifics of \solution\ in this current section.

\subsection{Handling Gradient Importance for the 2-stage Synchronization}

\solution\ reduces the communication burden during DDL training by deferring the transmission of specific insignificant gradients to ICS and utilizing an asynchronous gradient ranking algorithm to determine their importance. This section presents a mathematical demonstration of PGP and elaborates on the communication strategy between server and worker for GIB.

\subsubsection{Calculate the Gradient Importance}\label{sssection:gradient_importance}

The significance of parameters (or gradients) for each iteration can be approximated based on information from the previous iteration. As the synchronization of gradients for unimportant parameters has a limited impact on the next iteration's calculations, we propose evaluating the importance of parameters corresponding to a specific neuron.

In a neural network model, the criticality of a neuron can be determined by assessing the impact of its failure on the network's output, which can be achieved by setting the parameters of the corresponding neuron to zero and evaluating the entire model. While performing this test iteratively on all model parameters is not practical, mathematical formulas can provide a solution.

Given a neural network model with parameters $P = {P_0, P_1, P_2, ...P_n}$, and an input dataset $S$ with the loss function $L$, the objective of model optimization is to minimize $L(S, P)$.

Under the assumption that the parameters follow an identical independent distribution, the importance of a parameter ($D_k$) can be quantified as the difference in the loss with and without the specific parameter ($P_k$).

\begin{equation}
\begin{aligned}
D_k = (L(S, P) - L(S, P|_{P_k=0}))^2
\end{aligned}
\end{equation}

We perform a Taylor expansion on $D_k$. For simplification of computation, we adopt a first-order Taylor expansion in this case. However, higher precision can be achieved by using multi-order Taylor expansions.

\begin{equation}
\begin{aligned}
L(S,P) &= L(S,P|_{P_k=0}) + \frac{\partial L}{\partial P_k}(P_k-0) \\
& = L(S,P|_{P_k=0}) + \frac{\partial L}{\partial P_k}P_k
\end{aligned}
\end{equation}

Then, $D_k$ can be calculated by

\begin{equation}
\begin{aligned}
D_k &= (\frac{\partial L}{\partial P_k}P_k)^2 
&= (g_k P_k)^2
\end{aligned}
\end{equation}

The $D_k$ can be simplified to PGP to identify the importance of parameters or gradients. The importance of k-th parameter ($I_k$) can be represented as $I_k = |G_k P_k|$.

PGP can be utilized for determining the importance of gradients in the next iteration's RS, but evaluating the importance based on a single neuron unit would incur excessive computational complexities. To minimize these complexities, \solution\ calculates the importance $I$ on a per-layer basis. In other words, for a layer $l$, there exists

\begin{equation}
\begin{aligned}
I^l = \sum^j \lvert g_j P_j \rvert, \forall j \in l
\end{aligned}
\end{equation}

It is noteworthy that the work presented in~\cite{nvpgp-molchanov2019importance} reaches a similar conclusion. However, they do not extend the conclusion to distributed deep learning systems.

\subsubsection{Determine Size of Important Gradients} \label{sssection:important_gradient_size}

The next challenge is determining the number of layers of gradients that should be synchronized in the ICS, represented by $S(G^u)$. Since the computation time ($T_c$) of each iteration of DDL training is relatively fixed, we can provide a theoretical upper bound for the ICS size based on $T_c$, worker size $N$, and network attributes (bandwidth $b$, latency $l$, and loss rate $lr$).

For DL models with less than 1K layers, the GIB size would be smaller than 1KB. Thus, the synchronization time $T_{Push GIB}$ can be disregarded. Additionally, empirical experiments indicate the time consumption of $T_{Calc. GIB}$ is also relatively insignificant, which means

\begin{equation}
\begin{aligned}
& T_c \ge T_{ICS} + T_{Calc. GIB} + T_{Push GIB} \\
\Rightarrow & T_c \ge T_{ICS} \\
\Rightarrow & T_c \ge N \times S(G^u)/{b(1+lr)} \\
    \Rightarrow & S(G^u) \le b(1+lr) T_c / N = U_{max}
\end{aligned}
\end{equation}

The upper bound of $S(G^u)$ is denoted as $U_{max}$. Typically, a larger $S(G^u)$ can be chosen to decrease the communication overhead, thereby reducing the proportion of communication during training. However, a smaller $S(G^u)$ is beneficial for accuracy improvement at the beginning of the training. Thus, we design an algorithm that adjusts $S(G^u)$ to allow for a gradual increase to the upper bound until the end of the training. The loss of the ith epoch is denoted by $loss_i$, and it is assumed that when the training ends, the loss tends to 0. In the first training round, $S(G^u)_1$ is set to 0, and the $loss_1$ is set to $L$. In the subsequent ith training round, $S(G^u)_i = (1-\frac{loss_i}{L}) \times U{max}$.

As there is a linear correlation between the size of $T_c$ and the batch size, $S(G^u)$ may be equivalent to the size of all parameters in certain situations. A huge $S(G^u)$ can cause \solution\ to degenerate into an ASP. To mitigate these issues, we ensure that $U_{max}$ does not exceed 80\% of the model size. We use pseudo-code to represent the $S(G^u)$ tuning algorithm at Algorithm~\ref{code:gu_amount}.

\begin{algorithm}[H]
    \caption{$S(G^u)$ Tuning Algorithm}
    \label{code:gu_amount}
    \begin{algorithmic}[1]
      \STATE Set $U_{max}$ from network and other params
      \STATE $U_{max}$ = min($U_{max}$, 0.8 * ModelSize)
      \FOR{epoch $i$}
      \IF{Meet the end-of-training condition}
      \STATE EndTraining()
      \ENDIF
      \IF{$i == 1$}
      \STATE $L$ = $loss_1$
      \STATE $S(G^u) = 0$
      \ELSIF{$i > 1$}
      \STATE $S(G^u)_i = (1-\frac{loss_i}{L}) \times U{max}$
      \ENDIF
      \ENDFOR
    \end{algorithmic}
\end{algorithm}

\subsection{Preserving Training Accuracy} \label{sssubsection:preserve_accuracy}

In \solution, workers may experience training with stale parameters before the ICS is completed, which can negatively impact the final accuracy. \solution\ introduces LGP to overcome this issue. We assume that the k-th local gradient is $G_k^l$ and the global gradient is $G_k^g$, and use $G_k^l$ to update unimportant parameters at the end of RS. At the i-th iteration of training, we have Equation~\ref{equation:lgp}.

\begin{equation}
\begin{aligned}
P_{i, partial} = P_{i-1} + \sum^j{G_j^g} + \sum^k{G_k^g}, \\
\forall j \in G^i and\ k \in G^u
\label{equation:lgp}
\end{aligned}
\end{equation}

The $P_{partial}$ ensures that the training can at least use the local results of the previous iteration for the new iteration, which eliminates unnecessary training on these parameters.

\solution\ updates the parameters in semi-real time with the synchronized gradients from ICS. We assume that the gradient received by the worker for a fixed duration is $T$. At the end of this duration, \solution\ will immediately correct the parameters estimated locally by LGP to the global gradient, referring to Equation~\ref{equation:lgp_correct}.

\begin{equation}
\begin{aligned}
P_{partial} = P_{partial} - \sum^t{G_t^l} + \sum^t{G_t^g} \\ \forall t \in T
\label{equation:lgp_correct}
\end{aligned}
\end{equation}

Furthermore, the local dataset is shuffled every epoch for each worker to prevent a fixed portion of the dataset from always being trained with outdated parameters after LGP. 

We also propose another LGP method based on the Exponential Moving Average (EMA), called EMA-LGP, which uses an exponential average of the past global gradients and the current local gradients to perform the gradient correction for the current iteration. During evaluations, we find that EMA-LGP does not bring additional accuracy improvement but instead incurs more significant computational and memory overheads. The EMA-LGP is omitted in the \solution\, and we are conducting further experiments to verify its applicability.

\subsection{Degrading to ASP or BSP} 

\solution\ demonstrates excellent adaptability and versatility. The two-step synchronization design of \solution\ makes it robust and adaptable to cope with different DL models. Specifically, if all gradients are transmitted during RS, the synchronization model will be the same as BSP, and if all gradients are transmitted during ICS, it degrades to ASP. This degradation ensures the robustness of the \solution\ and guarantees that even if some models are incompatible with the 2-stage synchronization, the overall training process will still function correctly. 

\subsection{\solution\ with Co-located PS}

Existing PS architectures have an implementation that further utilizes the computational capabilities of worker nodes: co-located PS. In the \solution\ synchronization model, the working mode of co-located PS is easy to implement, but the potential impact on the computational capabilities of the corresponding worker(s) needs to be considered. Specifically, for the worker(s) selected as PS(es), the calculation of GIB needs to be carried out simultaneously during the FP and BP processes. These calculations include computing PGP, summing each layer of parameters, and sorting the sums of the layers. These steps can be performed on the GPU without parameter relocation to the CPU and have little impact on training. We have conducted a detailed evaluation of this overhead in \S~\ref{evaluation:overhead}.

\subsection{Implementation}

\begin{figure}[tbp]
\centering
\begin{minipage}[t]{\columnwidth}
\includegraphics[width=\columnwidth]{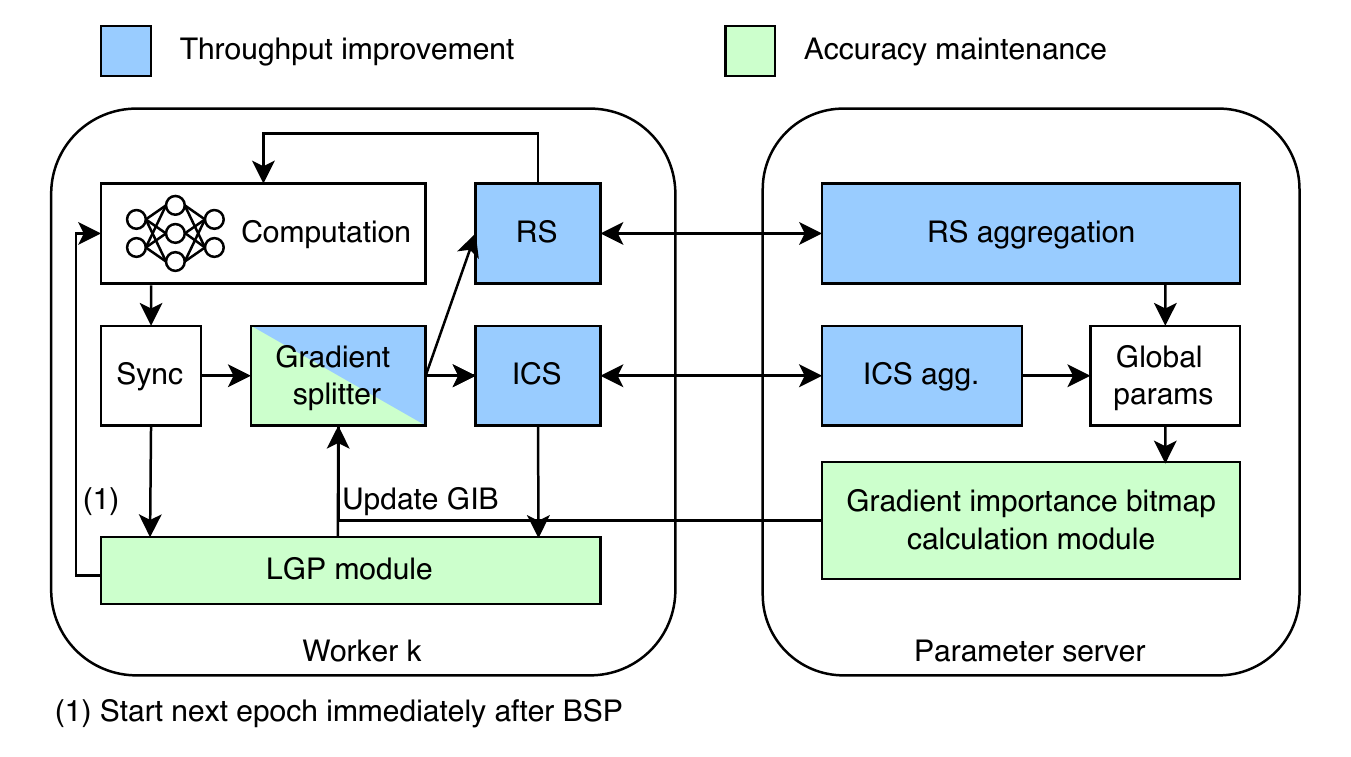}
\caption{\solution\ architecture.}
\vspace{-0.2in}
\label{figure:OSP_arch}
\end{minipage}
\end{figure}

We have implemented a prototype of \solution\ using PyTorch~\cite{paszke2019pytorch}. For efficient transmission of CUDA~\cite{cuda} tensor, we utilize the popular communication library NCCL as the communication backend. To circumvent the effects of global interpretation locks in python, we use processes instead of threads to parallelize computation and communication in ICS. However, as frequent process creation and destruction consume many system resources, we employ an efficient process pool to avoid this overhead. The architecture of \solution\ can be seen in Figure~\ref{figure:OSP_arch}. The modular prototype can be divided into two functions: throughput improvement and accuracy maintenance.

In the module for throughput improvement, we deploy the RS and ICS modules on both servers and workers for the synchronization actions of the two synchronization stages. At the worker side, after completing deep learning training backpropagation, the Gradient splitter module separates the local gradients to be synchronized according to the GIB received from the servers in the previous iteration into the important gradients $G^i$ and unimportant gradients $G^u$. As mentioned earlier, the following training iteration starts immediately after all workers finish RS and LGP. During the next iteration of training, ICS will be performed.

For the accuracy maintenance modules, the server uses an asynchronous GIB calculation module that works with the Gradient splitter on the worker side to mitigate the accuracy degradation caused by the split synchronous processes. On the worker side, the LGP module reduces the training accuracy degradation caused by training with stale parameters. 

\smallbreak

To summarize, \solution\ overlaps the communication and computation by splitting the synchronization processes of traditional BSP and ensuring that no additional computational overhead is introduced that affects training throughput. We discuss and resolve several challenges of \solution\ in this section, including gradient importance calculation, determining the amount of data for important gradients, and how to perform accuracy preservation. We conduct extensive experiments in the subsequent sections to evaluate the performance of \solution\ in terms of throughput improvement and accuracy preservation.

\input{Figures/PerformanceMetric}

\section{Evaluation} \label{section:evaluation}

In this section, we perform several experiments to evaluate the performance of \solution\ in terms of 1) training throughput improvement, 2) impact on convergence accuracy, and 3) additional computational overhead.

\subsection{Testbed Setup}

\subsubsection{Cluster Configurations}

In our experiments, we evaluate the performance of \solution\ using a 9-node cluster. Eight nodes serve as workers, while one node serves as the PS. All nodes are located in the same rack and interconnected through a HUAWEI CE6881-48S6CQ layer-3 top-of-rack (ToR) switch with 10Gbps links. Each node is equipped with 2 Intel(R) Xeon(R) Silver 4116 CPU @ 2.10GHz (24 cores, 48 threads), 256GB RAM, and one Intel(R) 10Gbps Ethernet Network Adapter X722. Additionally, each node has one NVIDIA Tesla T4 16GB GPU. The NVIDIA driver version is 460.91.03, and the CUDA~\cite{cuda} version is 11.2. The operating system is Ubuntu 18.04.2 with kernel version 4.15.0.

\subsubsection{Workloads}

In the evaluation, we conduct a series of experiments to evaluate the performance of various deep learning models on different benchmark datasets. We evaluate \solution\ on 5 different deep learning models with 4 popular datasets, which include training and testing ResNet50~\cite{resnet-he2016deep} and VGG16~\cite{vgg-simonyan2014very} models on the CIFAR10~\cite{cifar10-krizhevsky2009learning} dataset, InceptionV3~\cite{inceptionv3-xia2017inception} on the CIFAR100~\cite{cifar10-krizhevsky2009learning} dataset, ResNet101~\cite{resnet-he2016deep} on the ImageNet1K~\cite{deng2009imagenet} dataset, and the BERTbase~\cite{devlin2018bert} model on the SQUAD1.1~\cite{rajpurkar2016squad} dataset. The SQUAD1.1 dataset is used for the fine-tuning task on the BERTbase model. These evaluations are chosen to provide a comprehensive assessment of the models' capabilities across a range of image classification and natural language processing (NLP) tasks, thereby enabling a thorough understanding of the strengths and limitations of \solution.

\subsubsection{Baselines and Configurations}

We compare the performance of \solution\ against classical and state-of-the-art synchronization models, such as ASP, BSP, and R$^2$SP~\cite{r2sp-chen2019round}. Since PyTorch~\cite{paszke2019pytorch} does not have an official implementation for ASP, we implement it using NCCL's built-in efficient point-to-point communication primitive. The codes for BSP and R$^2$SP are taken from their official repositories. The initial learning rate is set to 0.1 and is halved every 10 epochs. The batch size for CIFAR-10, CIFAR-100, and ImageNet1K is 64, and for SQUAD1.1 is 12. All other hyperparameters are set to their default values. The models are recognized as converged when the accuracy does not greatly improve for 10 epochs.

\subsubsection{Performance Metrics}

We employ several metrics to evaluate the performance of the \solution. These metrics include:

\begin{enumerate}
    \item \textbf{Throughput.} It measures the number of processed samples (images or QAs) during a time period, which is an important indicator of the synchronization model's efficiency.
    \item \textbf{Top-1 accuracy.} This metric represents the maximum percentage of correctly classified samples during the whole training for image classification models. The F1 score is used in the NLP model (BERTbase), which is a harmonic mean of precision and recall, providing a single metric that balances both false positives and false negatives. These metrics are widely used to evaluate the model's convergence.
    \item \textbf{Number of iterations to top-1 accuracy.} This metric indicates the number of training iterations required for the model to achieve the desired top-1 accuracy. It is an important measure of the synchronization model for its impact on learning.
    \item \textbf{Batch Synchronization Time (BST).} BST quantifies the time taken for synchronizing the model parameters across different worker nodes during training. It is a key factor in understanding the scalability and efficiency of the synchronization models.
    \item \textbf{Time-to-accuracy curve.} This curve visualizes the relationship between training time and the achieved accuracy, allowing for the assessment of the model's learning progress over time. It provides insights into the model's training dynamics and can help visualize gaps in efficiency.
\end{enumerate}

By considering these diverse metrics, we aim to provide a comprehensive evaluation of the \solution, shedding light on its overall effectiveness, efficiency, and suitability for various training tasks. Besides these performance metrics, we further evaluate the computational overhead of PGP and ranking algorithms in \S~\ref{evaluation:overhead}. 

\input{Figures/TtA_cruve}

\subsection{Improvements on Performance Metrics}

We first use a complete round of training to evaluate the performance of the 4 methods - ASP, BSP, R$^2$SP, and OSP - in various performance metrics. To facilitate comparison, the unit for BERTbase is the number of Questions and Answers (QAs) trained per 10 seconds. In terms of throughput evaluation, \solution\ demonstrates superior performance compared to all other methods in image classification, while achieving near-ASP throughput in NLP tasks, as shown in Figure~\ref{figure:thru_improvement}. For the top-1 accuracy evaluation (Figure~\ref{figure:top1_acc}), \solution\ can reach near-optimal top-1 accuracy compared to the BSP and R$^2$SP, while ASP performs the worst. 

Another criterion for evaluating the impact of synchronous models on the entire training process is to compare the number of iterations required for the models to converge to optimal accuracy (or F1 accuracy). We present the complete evaluation results in Figure~\ref{figure:iterations}, where the BERTbase model uses 67 batches as one iteration for ease of presentation. Experiments show that the number of iterations for \solution\ compared to BSP does not significantly increase and may even decrease in some specific cases. We conjecture that this may be due to the potential of \solution\ to enhance the model's generalization. The comparison of BST (Figure~\ref{figure:bst}) shows that the synchronization time per round of \solution\ can be significantly reduced, which is the key to improving the training task throughput. This also ensures that in the worst case, even if \solution\ increases the number of training iterations, efficient convergence speed can still be guaranteed.

\subsection{Time-to-accuracy Curve Analysis}

We evaluate the impact of the throughput improvement of \solution\ on the model convergence accuracy by evaluating the time-to-accuracy curve, as shown in Figure~\ref{figure:TtA} and Figure~\ref{fig:bert-base}. As an integration of the aforementioned performance metrics, these figures can intuitively reflect the performance advantages and accuracy convergence of deep learning tasks using the \solution\ synchronization model. The figures indicate that \solution\ has no accuracy loss in model convergence, which means that the \solution's throughput advantage can be fully translated into a faster convergence speed. The advantage of convergence speed is greater in image processing tasks, and it also has certain advantages in NLP tasks. The curve is generally in line with the results of the performance metric.

\subsection{Overhead in Co-located PS Scenarios}\label{evaluation:overhead}

Since \solution\ involves additional computational overhead at the PS side, it does not affect the training of workers when the training task uses a standalone PS (standalone PS). However, the traditional PS architecture also supports using one of the worker nodes as a PS (co-located PS). We further assess the impact of using a worker as a PS node on the computational process. We perform a preliminary co-located PS deployment (\solution-C) on \solution: the selected PS worker begins training after completing PGP calculations and sorting. We compare the \textbf{batch computation time (BCT)} of workers in three scenarios: BSP (using standalone PS), \solution-S (using standalone PS), and \solution-C (using co-located PS), to evaluate the potential additional overhead introduced by \solution\ in co-located scenarios.

Figure~\ref{fig:overhead} is an experiment on the potential additional computational overhead caused by \solution. From the evaluation results, it can be seen that compared to BSP (standalone PS), OSP-S introduces almost no additional computational overhead for the worker. OSP-C leads to a limited additional computational overhead, with the lowest being the InceptionV3 model (3\%) and the highest being the VGG16 model (8\%). These additional overheads are acceptable and can be mitigated by increasing the scale of the cluster or batch size tuning. It is worth noting that we perform the evaluation on a preliminary implementation of OSP-S. If memory capacity and computational capability permit on the co-located PS, PGP calculation and sorting can be fully parallelized with the FP and BP processes in model training, in which the computational overhead of the \solution\ will further decrease.

\smallbreak

In summary, the \solution\ can achieve significant throughput advantages from the 2-stage synchronization. \solution\ leads among multiple performance metrics compared to the state-of-the-art synchronization models, attains little loss in accuracy, and has only a small additional computational overhead.

\section{Discussion} \label{section:discussion}

\subsection{Handling Scaling-up}

As the number of training nodes increases, relying on a single PS to synchronize global gradients can lead to prolonged communication times or even result in the \solution\ transmitting a large number of gradients during ICS. To address this challenge, as the number of nodes increases, the \solution\ can utilize multiple PS servers simultaneously for global gradient synchronization and divide independent \solution\ processes into different synchronization groups, similar to BytePS~\cite{byteps-jiang2020unified}. Determining how to orchestrate the PS(s) for the PGP calculation of the entire model and grouping parameters to assign different aggregated PS(s) are critical areas of our future work.

\subsection{Facing Heterogeneous Environments}

\solution\ also encounters challenges in heterogeneous environments. Typically, heterogeneity can be divided into computation capability and communication heterogeneity. Computation capability heterogeneity refers to variations in the computational capabilities of different worker nodes, while communication heterogeneity refers to variations in access bandwidth among different computing nodes. Referred to previous approaches like~\cite{r2sp-chen2019round}, computation capability heterogeneity can be effectively addressed by using batch-size tuning to ensure that all nodes have the same computation time. However, communication heterogeneity is a common challenge faced by existing synchronization models and is another potential research area for \solution.

\section{Related Work} \label{section:related_work}

As deep learning advances, models and datasets grow, rendering single-node training insufficient. Distributed training techniques are introduced for efficient large-model training. However, frequent node communication during distributed training increases communication overhead, limiting scale and efficiency. To reduce overhead and speed up training, research focuses on three aspects: synchronous models, model compression, and network optimization.


\textbf{Synchronous models.} The synchronous model term pertains to the timing and method of parameter update synchronization among nodes. In recent years, in addition to traditional BSP and ASP, novel synchronous models have been developed to expedite distributed training. To tackle the incast problem in BSP, R$^2$SP~\cite{r2sp-chen2019round} was introduced, scheduling communication among all nodes in a round-robin fashion. Despite addressing the Incast issue, R$^2$SP still faces challenges related to substantial communication overhead and stale gradients. DSSP~\cite{zhao2019dynamic} is an SSP variant capable of dynamically adjusting the threshold between the fastest and slowest nodes based on real-time processing speeds during model training. However, DSSP experiences performance degradation in heterogeneous environments. To accelerate model training in heterogeneous environments while maintaining accuracy, CASP~\cite{zhou2020petrel} employs clustering techniques to categorize nodes according to their performance. Within each group, nodes with similar performance utilize the BSP synchronous model for training, while nodes with significant performance disparities between groups adopt the ASP synchronous model. Although CASP demonstrates effectiveness in heterogeneous environments, its BSP mechanism still suffers from suboptimal bandwidth utilization. \solution\ enhances bandwidth utilization through parallel computation and communication, further addressing the limitations of existing approaches.

\textbf{Model compression.} Model compression is a crucial approach for reducing communication overhead in DDL~\cite{han2015deep}. To minimize communication costs, lots of model compression algorithms have emerged in recent years. The most commonly employed are Random-K~\cite{randk-stich2018sparsified} and Top-K~\cite{topk-aji2017sparse}. Random-K arbitrarily selects a subset of K\% parameters for transmission, whereas Top-K sorts all parameters and selects the top K\%. 
Additionally, some research efforts focus on representing parameters with fewer bits. For instance, 8-bit quantization~\cite{dettmers20158} converts model parameters from 32 bits to 8 bits to reduce the transmission volume. 
GRACE~\cite{xu2021grace} offers a comprehensive analysis and comparison of existing compression algorithms, providing valuable insights for future developments.

\textbf{Network optimization.} Communication is a bottleneck that limits the scale and efficiency of distributed training, and therefore network optimization is essential. As deep learning tasks are iterative and do not require precise calculations, losing a portion of the data during each iteration has little impact on the final model performance. Based on this, the paper~\cite{xia2019rethinking} calls on the community to study a set of packet loss tolerant transport protocols to accelerate DDL. DGT~\cite{zhou2021dgt} distinguishes the importance of gradients and then transmits important gradients using reliable TCP and unimportant gradients using UDP. In addition, programmable networks also provide opportunities to accelerate DDL. ATP~\cite{lao2021atp} and SwitchML~\cite{sapio2019scaling} leverage the powerful processing capabilities of programmable switches to aggregate parameters in the network, thereby reducing network traffic. These works are orthogonal to \solution\ and can be used complementarily.

\section{Conclusion} \label{section:conclusion}

In this paper, we propose \solution, a synchronization approach that enhances the throughput of PS-based DDL training by overlapping the communication with the computation. \solution\ employs a 2-stage synchronization strategy to boost training throughput by separating the synchronization into RS and ICS based on the real-time network situation. Additionally, \solution\ introduces a gradient splitting method that incurs no additional computational overhead based on PGP and uses LGP to maintain accuracy and avoid the stale parameter problem caused by delayed synchronization gradients. We have implemented \solution\ on PyTorch and conducted testbed evaluations. Experimental results demonstrate that \solution\ can achieve up to a 50\% throughput improvement over conventional and state-of-the-art gradient synchronization models without sacrificing final accuracy. Furthermore, the overhead associated with \solution\ has been confirmed to be within acceptable limits.

\begin{acks}

This work is sponsored by the Key-Area Research and Development Program of Guangdong Province (2021B0101400001), National Natural Science Foundation of China (62150610497, 62172108, 62002066), Natural Science Foundation of Shanghai (23ZR1404900), the Major Key Project of PCL, and Open Research Projects of Zhejiang Lab (2022QA0AB07). We also sincerely appreciate the anonymous reviewers for their valuable and constructive feedback.

\end{acks}

%% file: Figures/PerformanceMetric.tex

\begin{figure*}[tbp] 
\centering
\subfigure[Throughput]{\includegraphics[width=\columnwidth]{
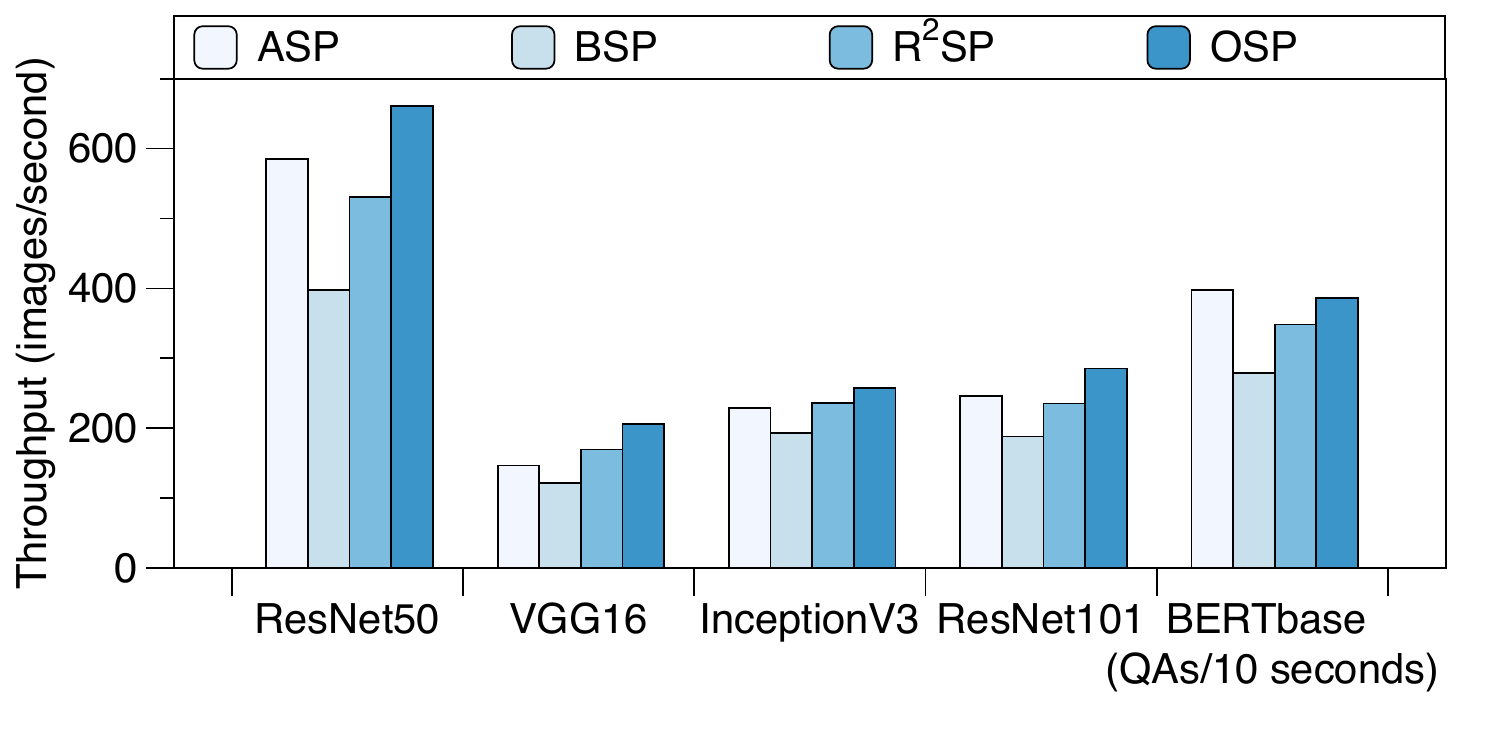}
\label{figure:thru_improvement}}
\subfigure[Top-1 accuracy.]{\includegraphics[width=\columnwidth]{
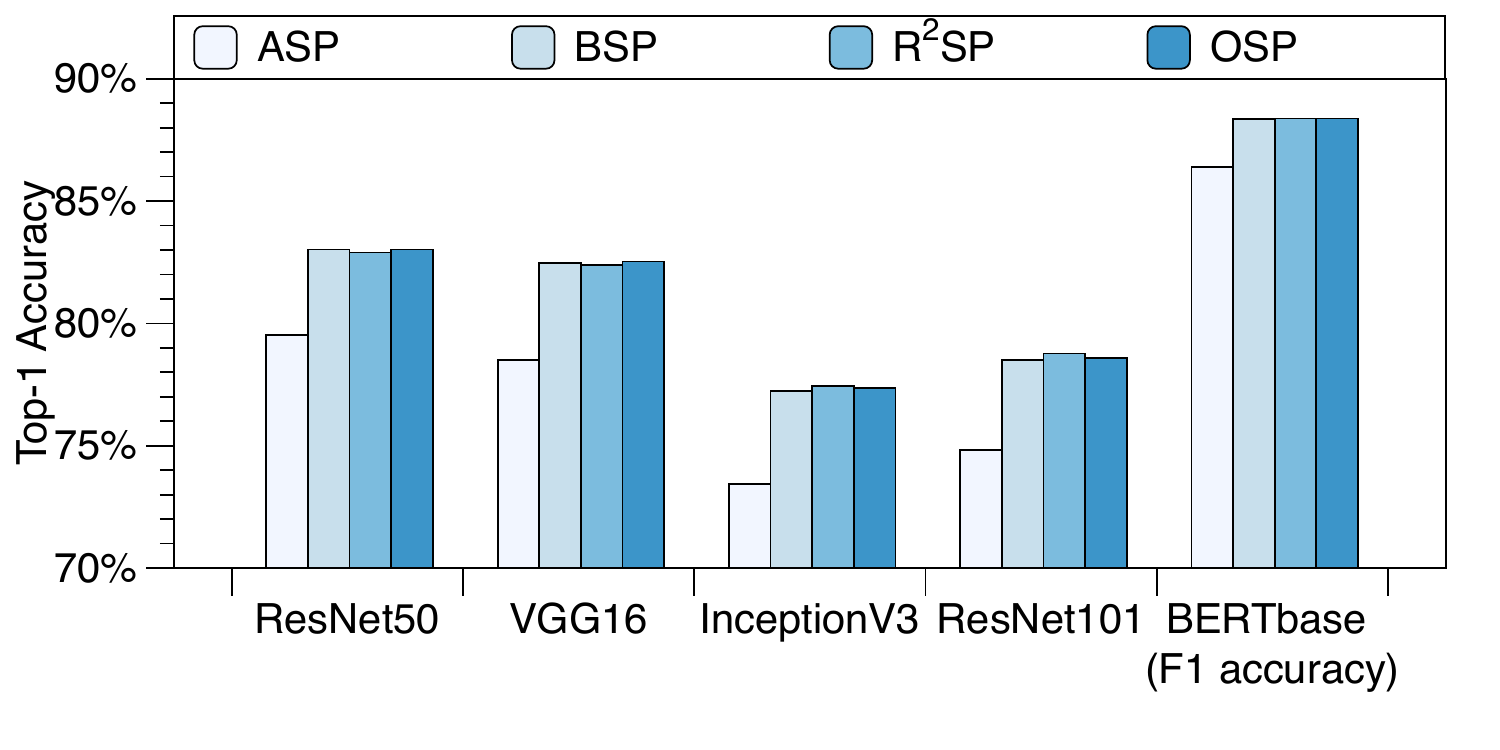}
\label{figure:top1_acc}}
\subfigure[Iterations to top-1 accuracy.]{\includegraphics[width=\columnwidth]{
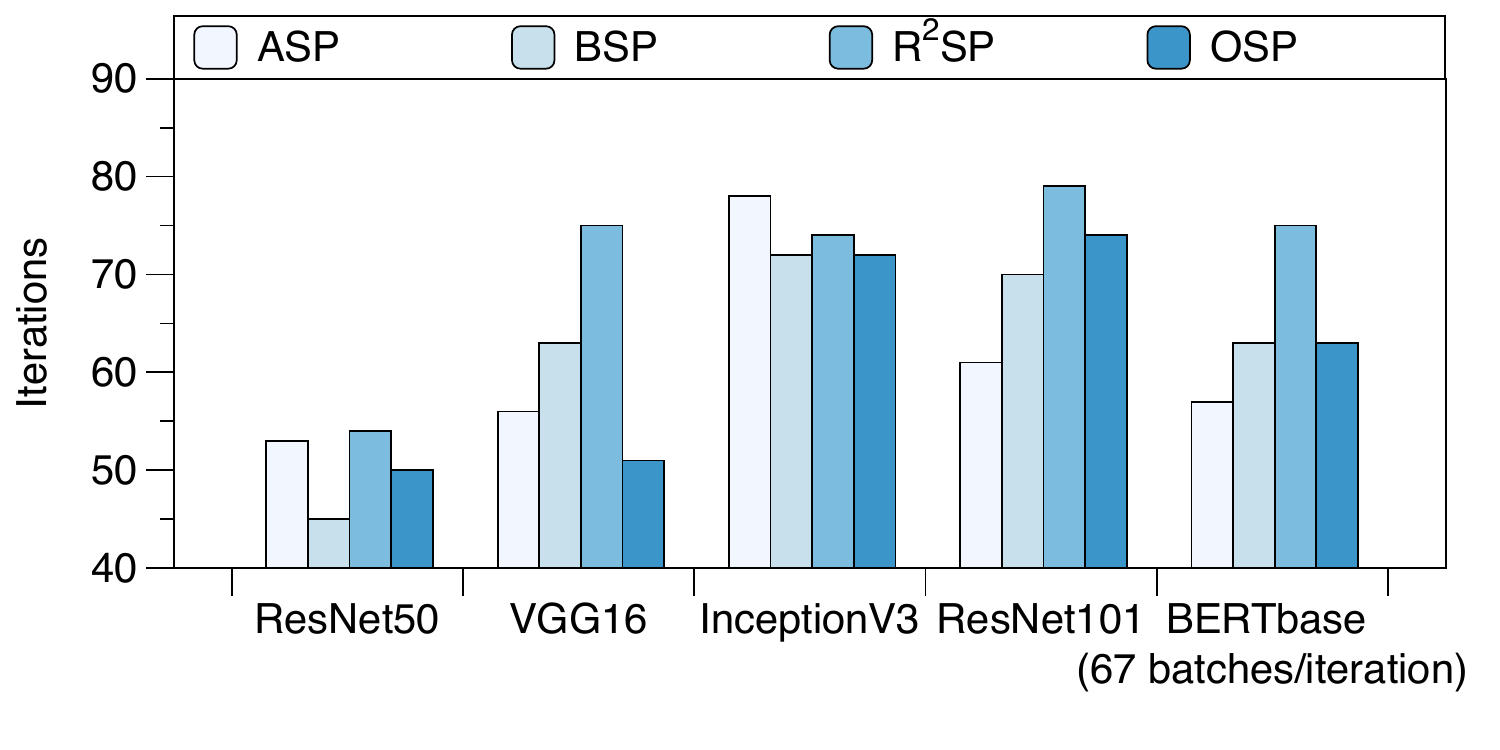}
\label{figure:iterations}}
\subfigure[Batch synchronization time.]{\includegraphics[width=\columnwidth]{
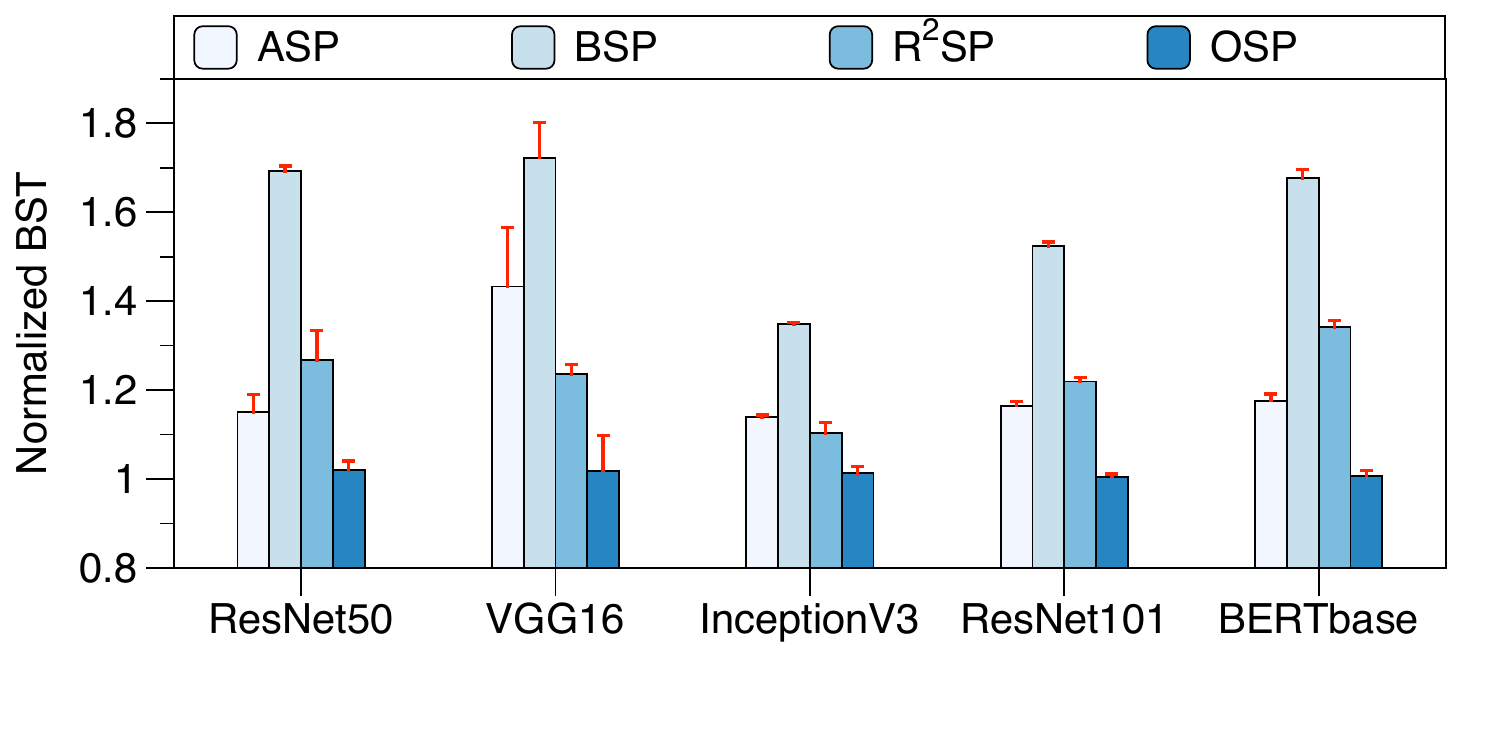}
\label{figure:bst}}

\caption{Evaluations on performance metrics.}
\vspace{-0.1in}
\label{figure:performance_metrics}
\end{figure*}

%% file: Figures/TtA_cruve.tex
\begin{figure*}[tbp]%
\centering
\begin{minipage}[th]{0.66\linewidth}%
\subfigure[ResNet50-CIFAR10]{\includegraphics[width=0.49\columnwidth]{
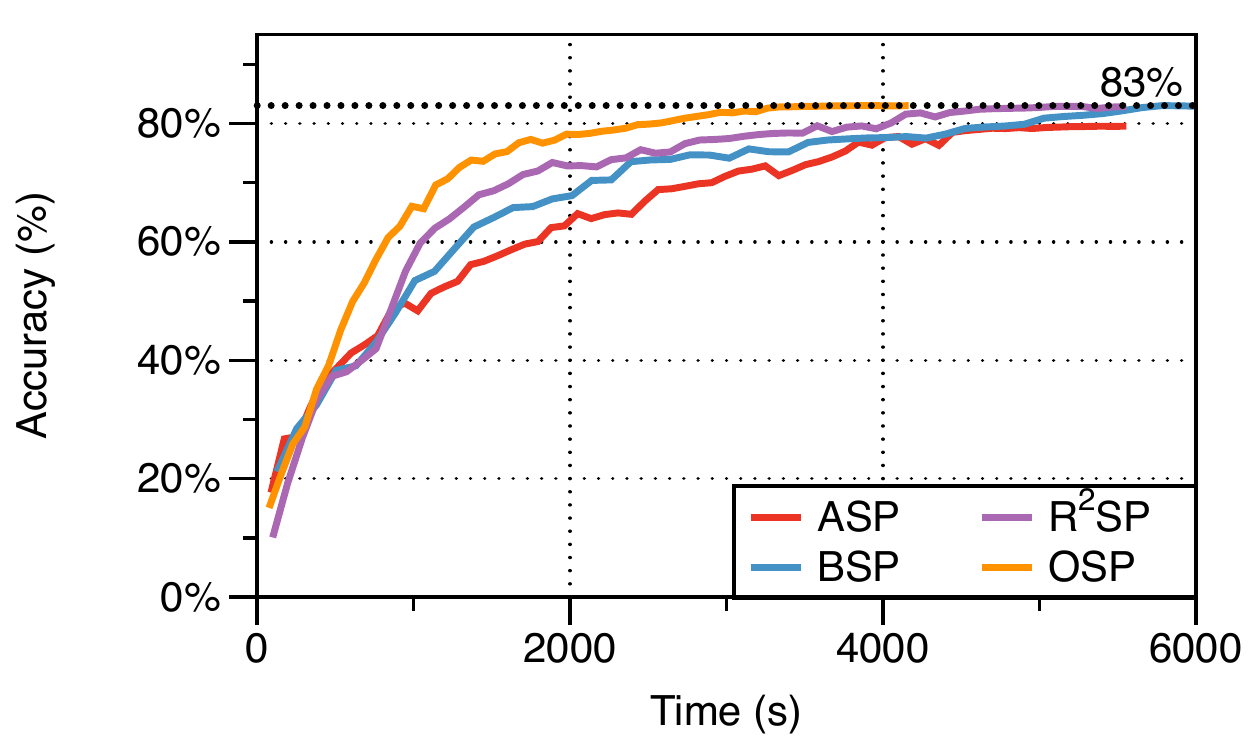}}%
\subfigure[VGG16-CIFAR10]{\includegraphics[width=0.49\columnwidth]{
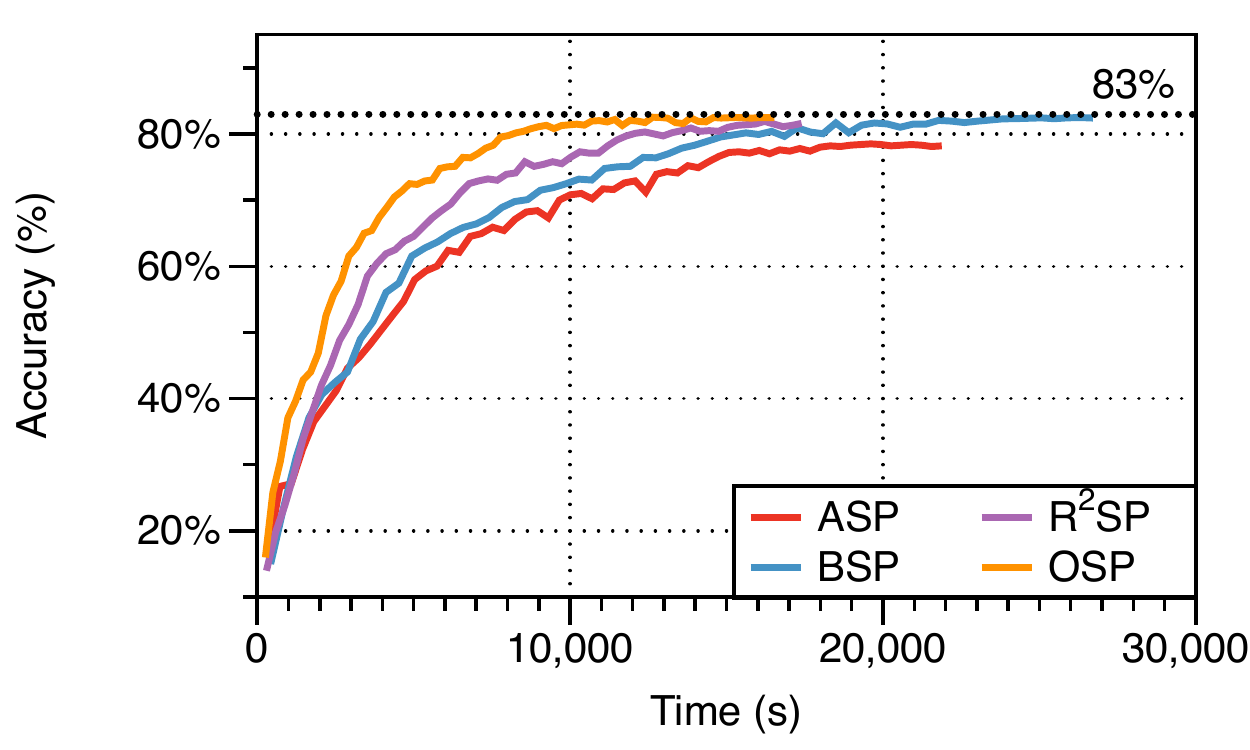}}
\subfigure[InceptionV3-CIFAR100]{\includegraphics[width=0.49\columnwidth]{
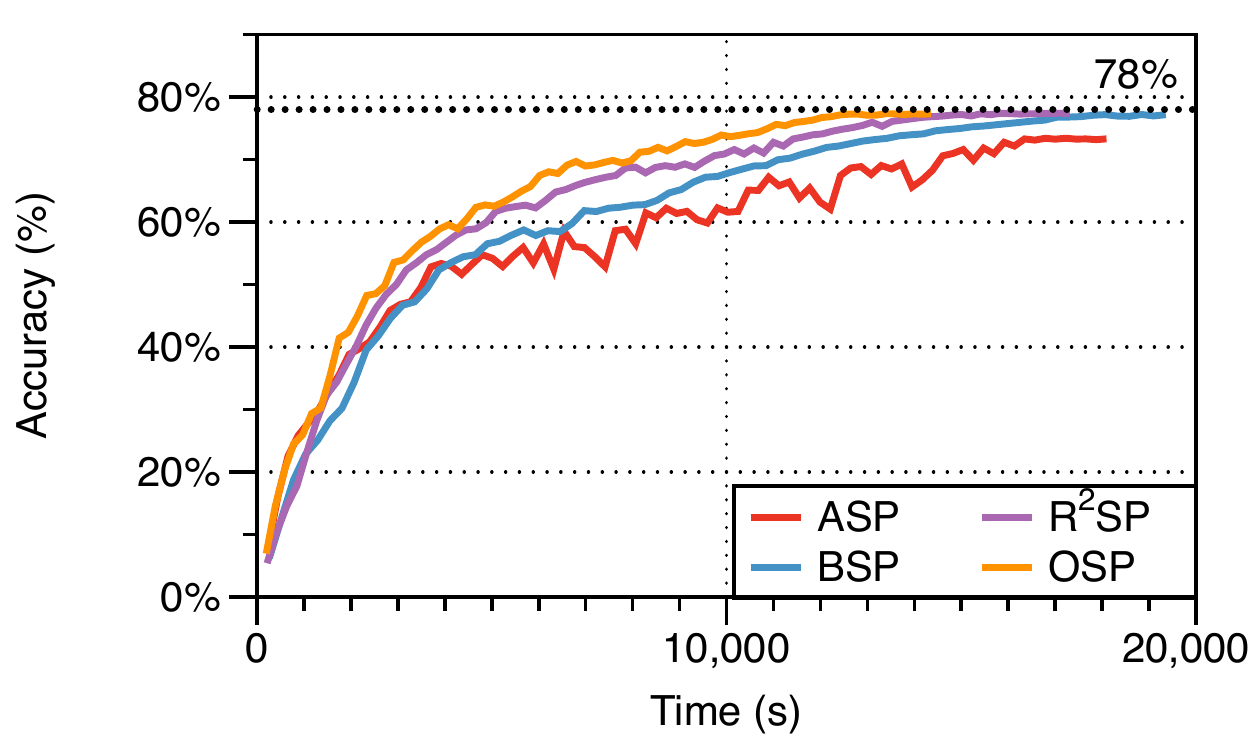}}%
\subfigure[ResNet101-ImageNet]{\includegraphics[width=0.49\columnwidth]{
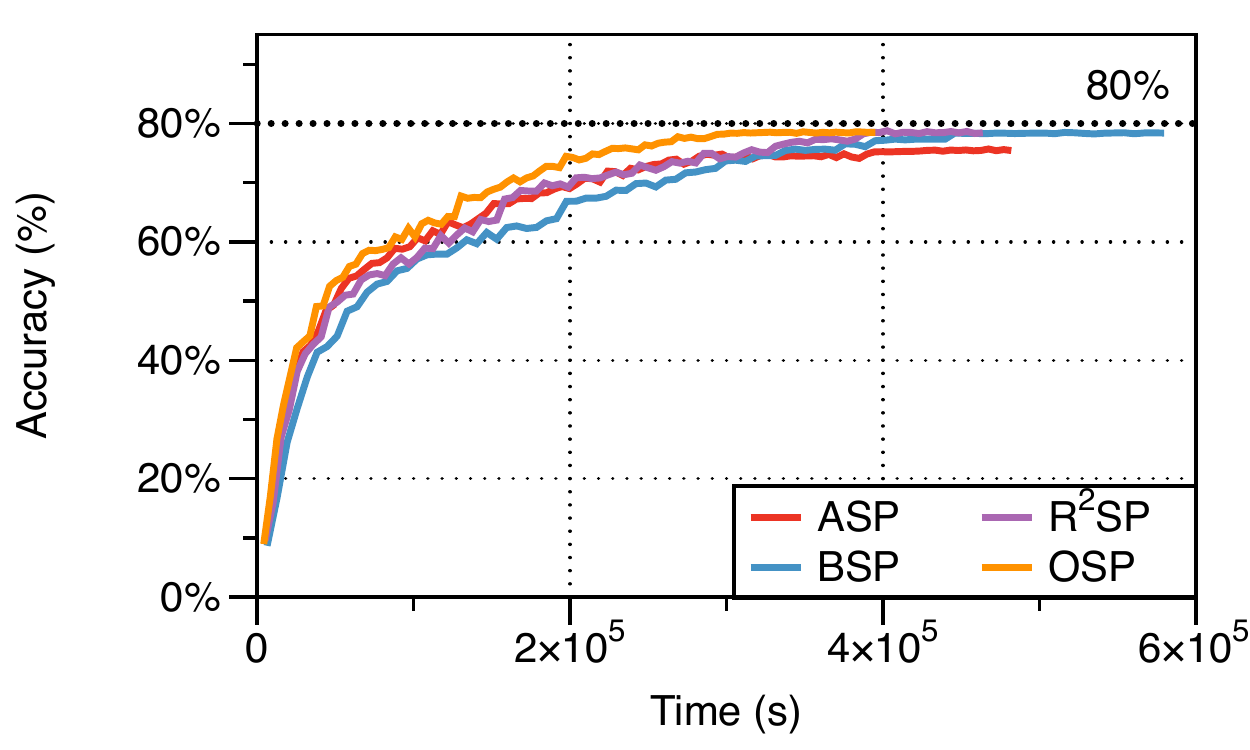}}
\caption{Time-to-accuracy curve on image classification tasks.}
\label{figure:TtA}
\end{minipage}%
\begin{minipage}[th]{0.33\linewidth}
\includegraphics[width=\columnwidth]{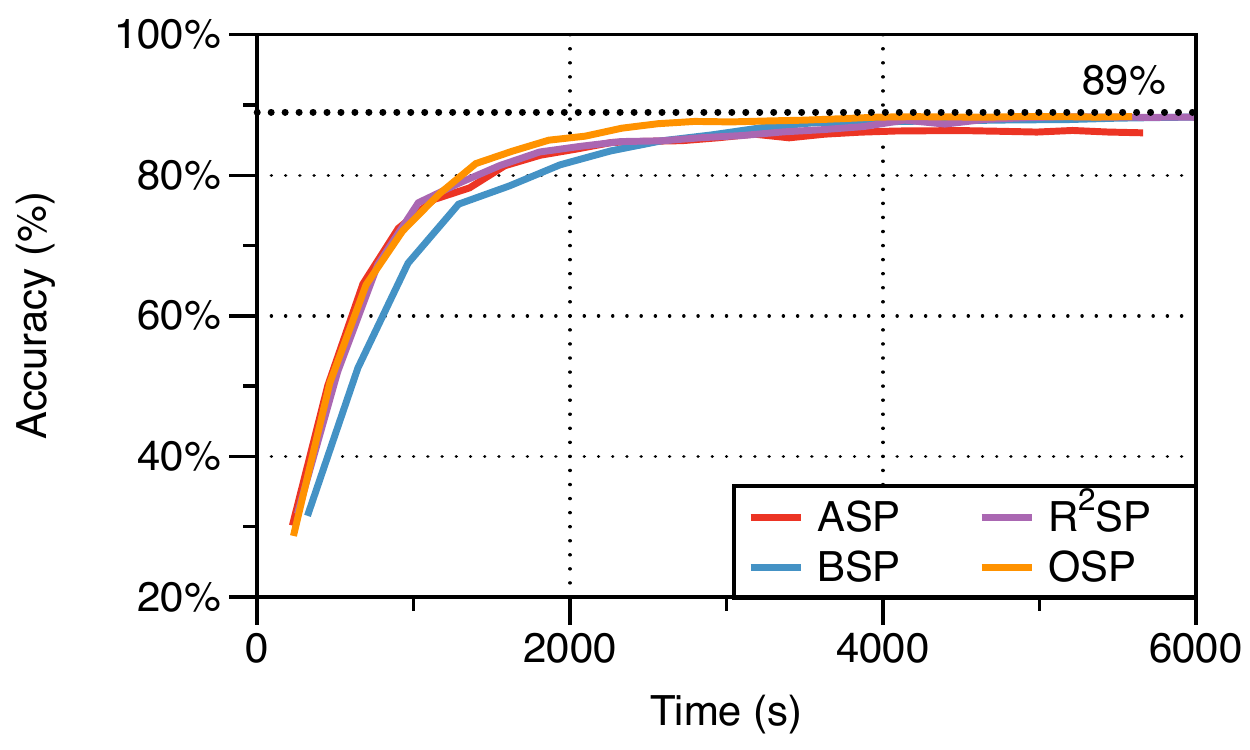}
\caption{Time-to-accuracy curve on the fine-tuning task of BERTbase with SQUAD1.1.}
\label{fig:bert-base}
\includegraphics[width=\columnwidth]{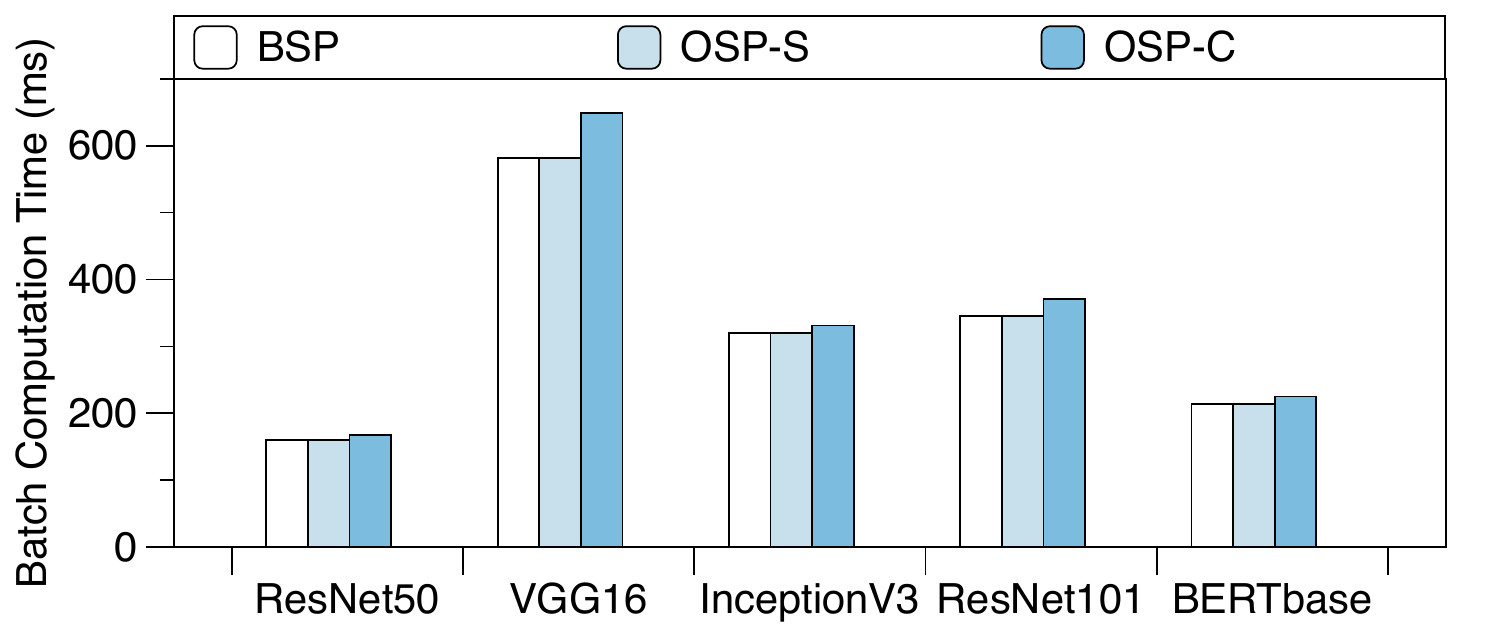}
\caption{Evaluations on the BCT of \solution\ with co-located PS.}
\label{fig:overhead}
\end{minipage}%
\vspace{-0.1in}
\end{figure*}
